\documentstyle[prl,aps,epsf]{revtex}
\begin{document}
\twocolumn[\hsize\textwidth\columnwidth\hsize\csname @twocolumnfalse\endcsname

\title{Towards the grain boundary phonon scattering problem:
an evidence for a low-temperature crossover
}
\author{V.A. Osipov and S.E. Krasavin}
\address{
Joint Institute for Nuclear Research,\\
Bogoliubov Laboratory of Theoretical Physics\\
141980 Dubna, Moscow region, Russia\\
}
\address{\em (\today)}
\preprint
\draft
\maketitle

\begin{abstract}
The problem of phonon scattering by grain boundaries is
studied within the wedge disclination dipole (WDD) model.
It is shown that a specific $q$-dependence of the phonon
mean free path for biaxial WDD results in a low-temperature
crossover of the thermal conductivity, $\kappa$.
The obtained results allow to explain
the experimentally observed deviation of $\kappa$ from a $T^3$
dependence below $0.1K$ in $LiF$ and $NaCl$.

\end{abstract}
\vskip 1cm

]

The effect of low-angle grain boundaries on the thermal
conductivity, $\kappa$, of $LiF$ and $NaCl$ over the temperature range
0.08--5 K has been investigated in~\cite{anderson,roth}. The main
conclusions are the following: (i) the boundaries are sessile,
(ii) the dominant phonon-scattering process comes from
static strain fields caused by boundaries, and (iii) the experimental
results are compatible with predictions of the theoretical
model~\cite{klemens} where a grain boundary is represented as a wall
of edge dislocations. In experiments, however,
in addition to the expected behavior $\kappa T^{-3}=const$,
a remarkable increase in $\kappa T^{-3}$
below $T^{*}\sim 0.1 K$ was detected.
A similar deviation, but beginning near $2 K$, was observed
in sapphire~\cite{wolf}.
There is still no satisfactory explanation of this phenomenon.
In particular, in~\cite{anderson} it was supposed that the measured
increase can be caused by the onset of partial specular reflection
from the lightly sandblasted walls.
Similarly, it was suggested that~\cite{wolf}
"a frequency independent scattering mechanism
should be present in these samples which becomes
ineffective below $1 K$".

In 1955 Klemens~\cite{klemens} studied the problem of
the scattering of lattice waves by grain boundaries
within the Born approximation.
Considering the grain boundary as an array of edge dislocations
lying in the plane of the boundary, he found that the phonon
mean free path is frequency independent.
Hence a $T^3$ dependence of the thermal conductivity
at low temperatures was associated with the boundary
scattering. While this finding explains well the experimental
results~\cite{anderson,roth,wolf} above some characteristic
temperature, $T^{*}$, it fails to describe the observed anomaly below
$T^{*}$.
It is important to note in this connection that
the result~\cite{klemens} was obtained under assumption that
the dislocation wall is {\it infinitely} long.
For a finite wall of well separated dislocations the problem
of the phonon scattering becomes difficult and is still unresolved.

An alternative model for description of grain boundaries
has been presented in~\cite{li1}. It was proposed that
grain boundaries being rather rotational than translational
defects can be described more naturally by disclinations.
Moreover, what is important, the far strain fields caused
by wedge disclination dipoles (WDD) were found to agree
with those from {\it finite} walls of edge
dislocations~\cite{li2,wit}. For this reason, the WDD-based model
allows us to study important effects due to finiteness
of grain boundaries.

Notice also that additional interest to this problem was inspired
by recent consideration of disclinations and dipoles of
disclinations in the context of metal glasses~\cite{nelson,kleman},
graphite films~\cite{tamura}, and nanostructures~\cite{chico}.
For example, an attractive model for a metallic glass
proposed in~\cite{nelson} results from disordering of a Frank-Kasper
disclination network. Graphite films are expected to contain a number
of disclination pairs due to the presence of the five- and the
seven-membered rings~\cite{tamura}.
Disclinations and disclination dipoles are
of importance in graphite nanotubes which have attracted
great interest recently.

At the same time, it is known~\cite{klemens} that information about
the nature of the principal imperfections in crystals can be extracted
from a study of the thermal conductivity.
In particular, the temperature dependence of the thermal conductivity
arising from defects is governed by the frequency dependence of
their scattering cross section for lattice waves.

In this Letter, we study the problem of phonon scattering
by grain boundaries in the framework of the WDD picture.
These scatterers are shown to possess rather specific
properties. In particular, the q-dependence of the phonon mean
free path varies significantly for different types of dipoles.
We show that the experimental results~\cite{anderson,roth} are
in good agreement with the calculated thermal conductivity due to
phonon scattering by {\it biaxial} WDD both above and below $T^{*}$.

Let us calculate a mean free path of phonons of frequency $\omega$
scattered by
the potential associated with a static deformation of a lattice
caused by straight WDD. Following the generally accepted
approach~\cite{klemens,ziman}
we consider an effective perturbation energy due to the strain field
caused by a single WDD in the form
\begin{equation}
\label{eq1}
U(\vec r) = \hbar\omega\gamma SpE_{AB},
\end{equation}
where $\hbar\omega$ is the phonon energy with the wavevector $\vec q$,
$\omega=qv_s$, $v_s$ is the sound velocity (it is assumed that
three acoustic branches are equivalent),
$\gamma$ is the Gr\"{u}neisen constant, and $E_{AB}$ is
the strain tensor due to WDD.

To simplify the problem, we assume that incident phonons are normal
to disclination lines. The suitable geometry is chosen:
disclination lines are directed along the $z$-axis,
the dipole's arm is oriented along the
$x$-axis. In this case, the strain matrix for WDD is known
(see, e.g.,~\cite{wit}), and  $U(\vec r)$ takes the form:
\begin{eqnarray}
\label{eq2}
U(x,y) &=& B\Biggl [\frac{1}{2}\ln {\frac{(x+L)^2+y^2}{(x-L)^2+y^2}}-
l_1\frac{x+L}{(x+L)^2+y^2} \nonumber \\
&+&l_2\frac{x-L}{(x-L)^2+y^2} \Biggr ],
\end{eqnarray}
where $B=\hbar qv_s\gamma\nu (1-2\sigma)/(1-\sigma)$,
$\nu $ is the Frank index, $\sigma$ is the Poisson constant,
$2L$ is the dipole separation, and
parameters $l_1$ and $l_2$ specify displacements of axes of rotation
from points where disclination lines pierce the $xy$ plane.
Notice that all possible types of WDD are included in Eq.(2).
In general, it describes a biaxial WDD with arbitrarily arranged
axes of rotation. For $l_1-l_2=2L$, $l_1=-l_2$, and $l_1=l_2=0$
one obtains the uniaxial WDD,
the symmetrical uniaxial WDD, and
the biaxial WDD with nonskew axes of rotation, respectively.

According to Eq.(2), at chosen geometry the problem reduces
to the two-dimensional case with
the phonon mean free path given by
\begin{equation}
\label{eq3}
\Lambda^{-1}_q = n_{i}\int_0^{2\pi}(1-\cos\theta)\Re(\theta)d\theta.
\end{equation}
Here $\Re(\theta)$ is an effective differential scattering radius,
and $n_i$ is the areal density of defects.
Within the Born approximation $\Re(\theta)$ is determined as~\cite{ziman}
\begin{equation}
\label{eq4}
\Re(\theta) =
\frac{qS^2}{2\pi\hbar^2v_s^2}\overline{|<\vec q|U(\vec r)|\vec q'>|^2},
\end{equation}
where all vectors are two-dimensional ones, $S$ is a projected area,
the bar denotes an averaging procedure over
$\alpha$ which defines an angle between $\vec p=\vec q - \vec q'$
and the $x$-axis. In other words, it means the averaging over randomly
oriented dipoles in the $xy$ plane.
Evidently, the problem reduces to the estimation of
the matrix element in Eq.(4) with the potential from Eq.(2).
For this purpose, it is convenient to use the polar
coordinates $(r,\phi)$
\begin{eqnarray}
\label{eq5}
U(p, \alpha) &=& <\vec q|U(\vec r)|\vec q'> \nonumber \\
&=&\frac{1}{S}\int d^2\vec r \exp[ipr\cos(\phi-\alpha)]U(r,\phi).
\end{eqnarray}
Here $p=|\vec p|=2q\sin(\theta/2)$.
Omitting the tedious calculations (the details will be
presented elsewhere) we write out the final result
\begin{eqnarray}
\label{eq6}
U(p, \alpha) &=&  B\Bigl [
-\frac{4\pi i}{p^2}\sin(pL\cos\alpha) \nonumber \\
&+&\frac{2\pi i\Delta_l}{p}\cos\alpha\cos(pLcos\alpha)\Bigr ],
\end{eqnarray}
where $\Delta_l=l_1-l_2$.
After averaging of $|U(p,\alpha)|^2$ over $\alpha$
and following integration in Eq.(3) with respect to $\theta$ one obtains
\begin{eqnarray}
\label{eq7}
\Lambda^{-1}_q &=& D^2(\nu L)^2n_iq\Biggl \{z^2\left (\frac{1}{2}+
J_0^2(2qL)\right) \nonumber \\
&+&\left (8-\frac{z(z+8)}{2}
\right)\left (J_0^2(2qL)+J_1^2(2qL)\right)   \nonumber \\
&-&\frac{4}{qL}J_0(2qL)J_1(2qL)\Biggr \},
\end{eqnarray}
where $D=\pi\gamma(1-2\sigma)/(1-\sigma)$, $z=\Delta_l/L$, and
$J_n(t)$ are the Bessel functions.
It should be emphasized that Eq.(7) is the exact result
which allows to describe all types of WDD.
Notice, that the behavior of $\Lambda_q$ in Eq.(7)
is actually governed by the only parameter $2L$ which characterizes
the dipole separation. Before proceeding to important applications
of Eq.(7) let us consider two limits to early solved problems.
As mentioned above, a biaxial WDD
(more precisely for $l_1=l_2=0$)
can be simulated by a finite wall of edge dislocations with
Burgers vectors situated in parallel.
As a consequence, with decreasing a dipole separation
such dipole should be equivalent in its
properties to a single edge dislocation with the Burgers vector
$\vec b=4\pi L\nu\vec e_y$.
Indeed, for small $L$ one can obtain from Eq.(7) that
$\Lambda\sim q^{-1}$. This is the well-known result for the phonon
scattering by edge dislocations. In the opposite limit of
large $L$, $\Lambda_q$ for a biaxial dipole becomes constant
(see details below) in agreement with the result obtained
for the infinite dislocation wall~\cite{klemens}.

In real materials, dipole separations which are of order of
the grain size can lie in the broad range between
$50$\AA\   (for nanocrystals) and $10^5$\AA\   (for polycrystals).
As it follows from Eq.(7), two specific
regimes of scattering appear depending on
the wavelength $\lambda$
of an incident phonon in comparison with $2L$.
As would be expected, the changing in behavior of $\Lambda_q$
occurs at $\lambda\sim 2L$. Thus, for thermal phonons,
we can estimate the typical
temperature of transition
\begin{equation}
\label{eq8}
T^{*} \approx \frac{\hbar v_s}{2Lk_B},
\end{equation}
where $k_B$ is the Boltzman constant.

An interesting consequence of Eq.(7) is the conclusion that
despite the seemingly similar nature of scatterers
the behavior of $\Lambda_q$ differs remarkably for different types
of dipoles. Let us examine two limiting cases.

1. Uniaxial dipoles, $\Delta_l=2L$ $(z=2)$. Eq.(7) takes the form
\begin{eqnarray}
\label{eq9}
\Lambda^{-1}_q &=& 2D^2(\nu L)^2n_iq\Bigl (1+
J_0^2(2qL)-J_1^2(2qL) \nonumber \\
&-&\frac{2}{qL}J_0(2qL)J_1(2qL)\Bigr ).
\end{eqnarray}
According to Eq.(9), for $qL\gg 1$ one obtains $\Lambda_q\sim q^{-1}$,
whereas for $qL\ll 1$ $\Lambda_q\sim q^{-5}$. Thus, in the case of
small wavelengths (in comparison with a dipole separation)
the phonon scattering due to uniaxial WDD behaves like
that for edge dislocations.
In the limit $\lambda\gg L$ the scattering of
phonons by uniaxial WDD is found to have a strong $q$ -dependence,
even stronger than the known Rayleigh scattering of phonons by point
impurities.
It should be noted that such behavior is compatible with an overall
view of uniaxial WDD as a strongly screened system~\cite{wit}.
It is worthwhile to mention that the low-angle uniaxial WDD can be
simulated by a finite wall of edge dislocations complemented
by two {\it additional} edge dislocations at both ends of the wall.
The sign of these dislocations is opposite to that of dislocations
in the wall and absolute values of Burgers vectors are equal to
$b=2L\tan (\pi\nu)$.
Notice that just these two dislocations provide a screening.
A significantly different picture arises in the case of the biaxial
WDD with $l_1=l_2$.

2. Biaxial dipoles with $\Delta_l=0$ $(z=0)$.
Eq.(7) transforms to
\begin{eqnarray}
\label{eq10}
\Lambda^{-1}_q &=& 8D^2(\nu L)^2n_iq\Bigl (J_0^2(2qL)+J_1^2(2qL) \nonumber \\
&-&\frac{1}{2qL}J_0(2qL)J_1(2qL)\Bigr ).
\end{eqnarray}
In this case, in the long wavelength limit one gets
$\Lambda_q\sim q^{-1}$ while for $\lambda< L$ we obtain that
$\Lambda_q\rightarrow const$.
It should be stressed that the appearance of the $q$-independent
region distinguishes remarkably this scatterer from
other types of WDD as well as from dislocations.

Fig.1 shows $\Lambda_q$ for three types of WDD
at $2q_DL=6\times 10^3$ (or $L\approx 2\times 10^3$\AA).
One can clearly see the characteristic points
where crossover occurs.
In accordance with our estimation in Eq.(8), crossover
takes place at $q/q_D\sim (1/6)\times 10^{-3}$ or, respectively,
at $T^{*}\sim \Theta/6000$
where $\Theta$ is the Debye temperature.
In the general case of arbitrarily arranged axes of rotation,
$0<\Delta<2L$, the mean free path interpolates
between two limiting curves.
The constant in Eq.(7) is chosen to be
$D^2(\nu L)^2n_iq_D=10^6cm^{-1}$.
\begin{figure}[h]
\epsfxsize=9cm
\centerline{\hspace{7mm} \epsffile{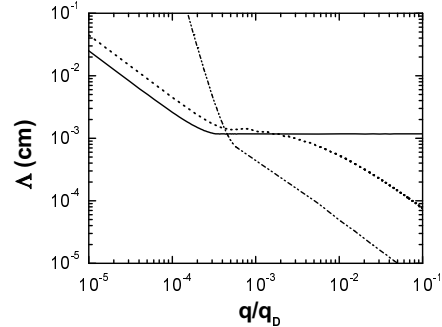}}
\vspace{3mm}
\caption{
Phonon mean free path $\Lambda_q$ as a function of
scaled frequency $q/q_D$ at $2q_DL=6\times 10^3$ for
$\Delta_l=0$ (solid line),
$\Delta_l=2L$ (dashed line), and
$\Delta_l=0.5L$ (dotted line).
The parameter set used is:
$L=1.35\times 10^{-5}cm$, $\nu=0.023$,
$D=2.6$, $n_i=1.8\times 10^7cm^{-2}$, and $v_s=4.8\times 10^5cm/sec$.
}
\label{fig1}
\end{figure}

Let us discuss briefly the applicability of the Born
approximation to the considered problem. Notice that,
as in the case of dislocations, the far field
of the potential in Eq.(2) behaves like $1/r$.
Thus, all arguments in support of the validity of the Born
approximation for the problem of phonon scattering
by dislocations (see, e.g.,~\cite{ziman,carrut}) are
appropriate here as well. Nevertheless, it is useful to
perform a rough estimation. In our case, by analogy with
unscreened Coulomb potential~\cite{stern}, we obtain
$2qL\ll (1-2\sigma)/[(1-\sigma)\nu\gamma]$. Thus, one can conclude
that the Born approximation should be valid
for description of low-angle WDD and/or small
dipole separation $2L$. Otherwise, one has to restrict the region
of admissible $q$.
One can expect, however, that the qualitative
behavior of $\Lambda_q$ will remain unchanged even beyond the
applicability of the Born approximation.
Notice also that when axes of WDD are orientated randomly,
one has to perform an additional averaging.
However, the known result for dislocations~\cite{klemens,ziman}
indicates that such averaging should lead only to a modification
of the numerical factor in Eq.(7).

Let us estimate a contribution to the thermal conductivity
caused by the phonon scattering due to WDD. For this purpose,
one can use the known kinetic formula written in the
dimensionless form
\begin{equation}
\label{eq11}
\kappa = \frac{k_B^4T^3}{2\pi^2\hbar^3v_s^2}
\int_0^{\Theta/T}x^4e^x(e^x-1)^{-2}\Lambda(x)dx,
\end{equation}
where $x=\hbar\omega/k_BT$, and the specific heat capacity
is chosen in the standard Debye form.
We have restricted ourselves to the thermal phonons with $T<\Theta$.
Then, for uniaxial WDD one obtains that
$\kappa\sim T^{-2}$ at low temperatures while
$\kappa\sim T^{-1}$ for $T\rightarrow \Theta$.

Let us examine Eq.(11) in detail for biaxial WDD with
$\Lambda_q$ from Eq.(10). In accordance with the above analysis,
the thermal conductivity should exhibit a crossover from
$\kappa\sim T^{2}$ to $\kappa\sim T^{3}$ with $T$ increasing.
Although the crossover temperature $T_0$ differs from $T^{*}$,
it is still determined by the value of $2L$.
\begin{figure}[h]
\epsfxsize=9cm
\centerline{\hspace{7mm} \epsffile{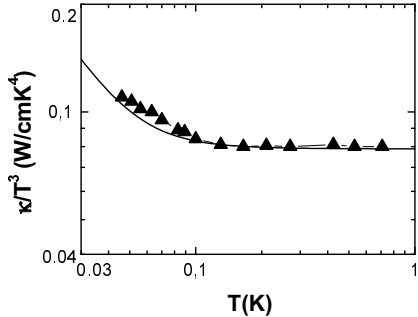}}
\vspace{3mm}
\caption{
Reduced thermal conductivity due to WDD scattering,
$\kappa\times T^{-3}$ vs temperature $T$,
calculated according to Eq.(11) with the same parameter
set as in Fig.1.
Measured points for the boundary-limited thermal conductivity
in $LiF$ (from Ref.[1]) are indicated by triangles.
}
\label{fig2}
\end{figure}
Fig.2 shows the reduced thermal conductivity calculated theoretically
according to Eq.(11).
We would like to stress that, within our scenario,
an increase in $\kappa T^{-3}$ below $T^{*}$
arises from the phonon scattering by biaxial WDD with $l_1=l_2$
and is quite universal. It should be observed in polycrystals
with appropriate sizes of grains.

We considered the experimental results for $LiF$~\cite{anderson,roth},
$NaCl$~\cite{roth}, and sapphire~\cite{wolf}.
Experimental data due to boundary-induced phonon scattering in $LiF$
from~\cite{anderson} are shown in Fig.2.
As is seen, the data are in a good agreement with our results for
appropriate choice of model parameters.
Notice that for a qualitative comparison we need
the only parameter which is a size of the grain.
For $NaCl$ the thermal conductivity was found to have
a similar behavior but with larger magnitude~\cite{roth}.
In sapphire, the measured crossover temperature
is essentially higher. This can be explained by smaller grain
sizes.
It is interesting to note in this connection that small $L$
can take place in glasses where the crossover temperature
$T_0\sim 1K$ and even higher. Thus, the dependence
$\kappa\sim T^{2}$ can be extended up to these temperatures.
A detailed study of physics of dielectric glasses on the basis
of biaxial WDD will be presented in a separate paper.

To conclude, the grain-boundary-induced phonon scattering has been
studied within the WDD-based model. The proposed model is shown to
take into account the finiteness of the boundary which is important
for thermal phonon scattering.
The phonon mean free path due to scattering by
static wedge disclination dipoles has been exactly calculated.
We have shown that the thermal conductivity exhibits a crossover
from $\kappa\sim T^{2}$ to $\kappa\sim T^{3}$ with $T$ increasing.
Thus, the biaxial WDD is a good candidate for
the specific scatterer proposed in~\cite{wolf}.
The results obtained allow us to explain the deviation
of the thermal conductivity from a $T^3$-dependence below $0.1K$
observed in $LiF$ and $NaCl$.
We expect that this crossover has a universal character
and should be observed in various materials
with the pronounced granular structure.

\vskip 0.5cm
\acknowledgments

We thank Sergei Sergeenkov for discussions and useful comments.
This work has been supported by the Russian Foundation for Basic
Research under grant No. 97-02-16623.

\end{document}